\documentstyle[12pt]{article}
\def\beq{\begin{equation}}   \def\eeq{\end{equation}}
\def\be{\begin{equation}}   \def\ee{\end{equation}}
\def\bea{\begin{eqnarray}}   \def\eea{\end{eqnarray}}

\begin{document}

\begin{flushright}
UND-HEP-01-BIG\hspace*{.2em}05\\
\today  \\ 
\end{flushright}
\vspace{.3cm}
\begin{center} \Large 
{MEASURING ABSOLUTE BRANCHING RATIOS OF CHARMED BARYONS IN 
$B$ DECAYS}\\
\end{center}
\vspace*{.3cm}
\begin{center} {\Large 
I. I. Bigi \\
\vspace{.4cm}
{\normalsize 
{\it Physics Dept.,
Univ. of Notre Dame du
Lac, Notre Dame, IN 46556, U.S.A.} }   
} \\
\vspace{.3cm}
e-mail address:\\
{\it bigi.1@nd.edu} 
\vspace*{1.4cm}

{\Large{\bf Abstract}}\\
\end{center} 
The $B$ factories are expected to provide  
huge samples of single $B$ decay events with little background by 
reconstructing one of the $B$ mesons produced in 
$\Upsilon (4S)$ decays. This represents a new experimental paradigm:  
such samples will allow to make measurements of a quality previously 
thought unrealistic. As example we discuss  
how absolute branching ratios for exclusive as well as 
inclusive charm baryon decays can
be extracted. One starts out by observing decays like 
$B^- \to \bar p X$ as a signature for 
$B^- \to \Lambda _c \bar p X$ etc. and then exploits various 
correlations of the flavour of the $B$ meson with the 
baryon number of the (anti)proton and other observables like 
the charge of a lepton, baryon number of another baryon etc.  
An integrated luminosity of about 500 $fb^{-1}$ as 
could be available by 2005 should be sufficient for the task.

\vspace*{.2cm}
\vfill
\noindent
\vskip 5mm

\tableofcontents

\section{The goal}

On the map for weak decays of charm baryons there are still 
large regions of terra incognita, which one wants to explore 
for two reasons. 
One needs {\em absolute} branching ratios for exclusive and 
inclusive decays of charm baryons as an engineering input 
for other studies concerning $B$ decays and their charm 
content, production rates etc.

Secondly,  
heavy quark expansions (HQE) have been developed into a mature 
theoretical technology for treating decays of heavy 
flavour  
hadrons. Employing the operator product expansion one expresses 
inclusive observables like lifetimes and total semileptonic 
widths in inverse powers of the heavy quark mass. Of course for 
charm decays one cannot count on more than a semiquantitative 
description. 
The weak decays of charm {\em baryons} provide a very 
rich phenomenology for probing HQE: beyond  
the lifetimes of $\Xi_c^+$ as well as $\Xi_c^0$ -- and preferably 
also of $\Omega_c$ -- one would like to measure 
also the inclusive
semileptonic  branching ratios of the charm baryons; this will be 
explained below.

In this short note we sketch the situation with inclusive 
charm baryon decays in Sect.\ref{BARINCL}, describe the 
method in Sect.\ref{METHOD} and present numerical 
estimates in Sect.\ref{NUM}, before summarizing in Sect.\ref{SUM}.

\section{Inclusive decays of charm baryons}
\label{BARINCL}

The $D^+-D^0$ and the $D_s-D^0$ 
lifetime ratios can readily be accommodated \cite{RIO}. Preliminary 
results from FOCUS and CLEO show, however, the 
$\Xi_c^+ -\Lambda _c^+$ lifetime ratio to be significantly 
larger than predicted: 
\beq 
\frac{\tau (\Xi_c^+)}{\tau (\Lambda _c)} = 
\left\{ 
\begin{array}{ll}
\sim 1.6 & {\rm quark\; model}\; \cite{BRANCO} \\ 
\sim 1.3 & {\rm HQE + quark\; model} \; \cite{MARBELLA}\\
2.8\pm 0.3 & {\rm CLEO} \; \cite{CLEO1} \\
2.29 \pm 0.14 & {\rm FOCUS} \; \cite{FOCUS} 
\end{array}
\right. 
\label{LRATIO}
\eeq 
This discrepancy could signal the inadequacy of the quark models 
used to evaluate the expectation values of the four-quark 
operators that enter in order $1/m_c^3$ \cite{VOL1}; or it could point 
to limitations in (quark-hadron) duality at the charm scale 
\cite{VADE1,VADE2}.

One can expect that the HQE yields a more reliable description 
for the {\em semi}leptonic widths of charm hadrons: there are fewer
contributions, and duality can be expected to provide a 
better approximation here than in nonleptonic 
transitions. Thus there are fewer excuses left for theorists here 
(although they can still come up with one).

Due to isospin invariance 
$  
\Gamma _{SL}(B^+) = \Gamma _{SL}(B_d) +{\cal O}(|V(ub)/V(cb)|^2)$,   
$\Gamma _{SL}(D^+) = \Gamma _{SL}(D^0) +{\cal O}(|V(cd)/V(cs)|^2)$;   
the ratio of the semileptonic branching 
ratios for these mesons therefore 
has to reflect their lifetime ratio. 
A priori there could be 
significant $SU(3)_{Fl}$ violations in 
$\Gamma_{SL}(B_s)$ vs. $\Gamma_{SL}(B_d)$ and 
$\Gamma_{SL}(D_s)$ vs. $\Gamma_{SL}(D^0)$; yet the HQE 
tells us that $SU(3)_{Fl}$ represents a good symmetry 
for these mesonic widths \cite{WA}.

No such argument can be made for the  
semileptonic widths of the baryons beyond 
$\Gamma _{SL}(\Xi^+_c) = \Gamma _{SL}(\Xi^0_c)$. 
On the contrary,
one expects large differences in the semileptonic 
widths of charmed baryons; e.g.,  
\beq 
\Gamma _{SL}(\Xi_c) \sim 2 \cdot \Gamma _{SL}(\Lambda_c)
\label{SL} 
\eeq 
through order $1/m_c^3$ \cite{VOLOSHIN}. 
This enhancement is due mainly to a {\em constructive} 
interference of the decay $s$ quark with the $s$ quark in the 
$\Xi_c$ wavefunction. Taking Eqs.(\ref{LRATIO}) and 
(\ref{SL}) together suggests that BR$_{SL}(\Xi_c^+)$ could 
be about five times larger than BR$_{SL}(\Lambda_c^+)$! Of course, 
BR$_{SL}(\Xi_c^+)$/BR$_{SL}(\Xi_c^0)$ $\simeq$ 
$\tau (\Xi_c^+)/\tau (\Xi_c^0)$ $\sim 3$ 
due to isospin invariance. These expectations can 
be summarized as follows: 
\beq 
{\rm BR}_{SL}(\Xi_c^+) \gg {\rm BR}_{SL}(\Lambda_c^+) \sim 
{\rm BR}_{SL}(\Xi_c^0) 
\label{BRSLEX} 
\eeq

It is highly desirable to find out 
whether such a dramatic effect exists. As indicated above, 
one has a simpler and more stable theoretical situation in 
{\em inclusive}  
{\em semi}leptonic decays. 
It would 
provide information on the baryonic expectation value of  
four-quark operators that affect also the 
$\Lambda _c$, $\Xi_c^+$ and 
$\Xi_c^0$ lifetimes. Invoking heavy quark symmetry they can be 
extrapolated to corresponding expectation values in 
the beauty sector \cite{VOL1}, where they affect the lifetimes 
of $\Lambda _b$ and $\Xi _b^{0,-}$ and the endpoint spectrum 
in semileptonic $B$ decays \cite{WA}.

It should also teach us lessons about the validity of duality 
at the charm scale. 
\footnote{It has been noted that roughly a third of the observed 
value of $\Gamma _{SL}(D)$ remains unaccounted for in the 
HQE result through order $1/m_c^3$. If that is an actual deficit, 
it might have its origin in a systematic underestimate of the 
charm quark mass $m_c$, which would have practically 
the same weight in all semileptonic charm widths; or it could 
be due to large {\em non}-factorizable contributions, 
which presumably would affect 
the various charm hadrons differently.}  
Beyond the intellectual value of such 
lessons, they  
could help us in properly interpreting $D^0 - \bar D^0$ 
oscillations \cite{DDBAR} as well as treating 
$B \to l \nu D^*$ \cite{VADE2}.

Since
contributions of  order $1/m_c^4$ can be quite sizeable, the factor of
two in  Eq.(\ref{SL}) has to be taken with quite a grain of salt. 
Accordingly one is not necessarily asking for a precise measurement
here.

\section{The method}
\label{METHOD}

The ideal set-up for measuring {\em absolute} branching ratios 
for exclusive channels and for inclusive transitions would be to
employ tagged events in 
$e^+e^- \to \Lambda _c \bar \Lambda _c$ and 
$e^+e^- \to \Xi _c \bar \Xi _c$. There are plans to create a 
tau-charm factory at Cornell; yet those plans do not envision 
to reach the $\Xi_c$ production threshold. The best value of 
BR$(\Lambda _c \to p K^- \pi ^+)$ has been inferred from 
continuum charm production $e^+e^- \to \Lambda _c X$ 
\cite{CLEO2}; yet the semileptonic branching ratio could 
not be obtained in such an environment, since the production rates for
the  various charm baryons are not known independently.

An alternative method is proposed here 
based on analyzing charm baryon 
production in $B$ decays, which
could be utilised at the  BELLE and BABAR beauty factories. 
The past success
of BELLE and  BABAR gives confidence that large data sets 
can be accumulated in $e^+e^- \to B \bar B$, where one of the 
mesons is fully reconstructed. Such a scenario represents 
a new paradigm in beauty physics: one can then 
envision to undertake
measurements  that before had not been viewed as feasible.  
The case
of charm baryon branching ratios discussed here is just one 
example for
this paradigm.

The basic method consists of three steps: first one reconstructs one 
of the $B$ mesons, which reveals the flavour of the other 
meson and at the same time reduces the number of tracks one 
has to contend with; then one identifies an (anti)proton among the 
remaining tracks to enrich the sample in charm baryons; finally 
one searches for one or more other particles that define the 
decay mode of the charm baryon one wants to study. A crucial tool 
here is the use of correlations between the $B$ flavour, the 
baryon number of the (anti)proton, the charge of the lepton etc.

It is quite conceivable that flavour tagging with only 
{\em partial} reconstruction of the first $B$ meson might 
suffice despite the presumably lower purity of the sample and the 
higher combinatorial background for the recoil $B$ meson, since 
the relevant correlations remain intact. Only detailed 
experimental studies can answer this question.

To give a more explicit description we 
focus on charged $B$ mesons for simplicity: 
$ e^+ e^- \to B^+ B^- $. 
Let us assume the $B^+$ has been reconstructed; then one 
knows that the remaining tracks have to belong to the other 
$B$ which we call the `recoil' $B$; its flavour is known 
as that of a $B^-$ from 
the reconstructed $B^+$. Next one searches for decays of the 
recoil $B^-$ into final states containing an antiproton --  
$B^-_{recoil} \to \bar p X^0$ -- which tells us that this final state 
has to contain a baryon as well; with 
$|V(cb)| \gg |V(ub)|$ there are three classes of 
such decays:

(i)  
$B^- \to \bar p + \Lambda_c^+ + X$;
 
(ii)   
$B^- \to \bar p + \Xi_c  K + X$;

(iii) 
$B^- \to \bar p + p/n D + X$.

\noindent  
Since $\Xi_c$ production requires the excitation of an 
$s\bar s$ rather than a $q \bar q$ pair, class (ii) will be reduced 
relative to class (i) by a factor of roughly three: 
\beq 
{\rm BR}(B \to \Xi _c^{+,0} X) \sim 
1/3 \cdot {\rm BR}(B \to \Lambda _c^+ X) 
\eeq 
The background class (iii) will be likewise reduced relative to 
class (i); even more importantly, its rate can be determined 
by observing the $D$ decays utilising known branching ratios.

To bias the sample towards $\Xi_c$ production, one selects 
$B^-_{recoil} \to \bar p K^+/\bar \Lambda + X^-$ requiring the 
correlation between the flavour of the $B$ meson and the observed 
baryon number and strangeness of the final state. 
Again, there are 
several classes of such decays:

(iv) 
$B^- \to \bar pK^+/\bar \Lambda  + \Xi_c   + X$;

(v) 
$B^- \to \bar pK^+/\bar \Lambda  + \Lambda_c \bar K   + X$;

(vi)  
$B^- \to \bar p K^+/\bar \Lambda + p/n D_s/D \bar K + X$

\noindent  
The background class (iii) can be controlled by observing 
the $D_{(s)}$ decays. 
The sample will still 
contain a roughly equal amount of $\Xi_c$ baryons and 
$\Lambda_c K$ combinations. We will return to the problem of 
controlling the latter, which represents a background here.

\subsection{Exclusive branching ratios}
 
First one conducts a validation or   
calibration measurement, namely 
extract BR$(\Lambda_c^+ \to p K^-\pi^+)$ by using the 
following ratio
\beq 
\frac{\Gamma(B^-_{recoil} \to \bar p (pK^-\pi^+)_{\Lambda_c} X)}
{\Gamma(B^-_{recoil} \to \bar p  X)} \simeq 
{\rm BR}(\Lambda_c \to p K^-\pi^+) 
\eeq 
and comparing the result with what is known 
from other measurements (and might be known even better in the 
future from data taken at a tau-charm factory). This will tell us 
to which degree one indeed controls the background and can 
identify $\Gamma (B^- \to \bar p +X)$ with 
$\Gamma (B^- \to \Lambda _c^+\bar p +X)$.
 
If this cross check works out satisfactorily, then one can turn to 
the more ambitious task of deducing 
$\Xi _c$ branching ratios by analyzing the sample
of 
$B^-_{recoil} \to \bar p K^+/\bar \Lambda + X^-$ events. The 
background due to class (vi) can, as already stated, be 
controlled with the observation of $D_{(s)}$ decays. 
Class (v) can be determined by measuring 
$B^-_{recoil} \to \bar p K^+/\bar \Lambda + 
(pK^-\pi^+)_{\Lambda_c} + X^-$ using 
BR$(\Lambda _c \to p K^- \pi^+)$. Let us call the remaining rate 
the `reduced' width; then one can infer: 
\beq 
\frac{\Gamma (B^-_{recoil} \to \bar p K^+/\bar \Lambda 
+ (f)_{\Xi_c} +X)}
{\Gamma ^{reduced}
(B^-_{recoil} \to \bar p K^+/\bar \Lambda +X)} \simeq 
{\rm BR}(\Xi_c \to  f) 
\eeq  
for a final state $f$ like $\Xi \pi \pi$ etc.

\subsection{Inclusive branching ratios} 

One can 
look for a (relatively soft) lepton with positive charge 
coming from  a semileptonic $\Lambda_c$ decay. Then one has 
\beq 
\frac{\Gamma(B^-_{recoil} \to \bar p l^+ X)}
{\Gamma(B^-_{recoil} \to \bar p  X)} \simeq 
{\rm BR}_{SL}(\Lambda _c) 
\eeq 
The strength of this method is that it employs a highly 
nontrivial correlation between the flavour of the recoil 
$B$ (inferred from the reconstructed $B$), the baryon 
number of the antiproton and the charge of the lepton.

An even more ambitious enterprise is to measure 
$\Gamma _{SL}(\Xi_c)$. Since the semileptonic width has to 
be the same for $\Xi_c^0$ and $\Xi_c^+$, yet $\Xi_c^0$ has a 
much shorter lifetime, measuring the 
larger BR$_{SL}(\Xi_c^+)$ represents a 
more favourable challenge. One can study the `reduced' width as 
defined above for 
$B_{recoil}\to \bar \Lambda X^-$ or 
$B_{recoil}\to \bar p K^+ X^-$ as a tag for 
$B_{recoil}\to \bar \Lambda/\bar pK^+ + \Xi _c^+ +X^-$ and then again 
searches for (relatively soft) positive leptons: 
\beq 
\frac{\Gamma(B^-_{recoil} \to \bar \Lambda/
\bar p K^+ l^+ X)}
{\Gamma(B^-_{recoil} \to \bar \Lambda/\bar p K^+  X)} \simeq 
{\rm BR}_{SL}(\Xi^+ _c) 
\eeq 
Here one has to exploit the correlation between the flavour of 
the recoil $B$, the baryon number of $\bar \Lambda $ or $\bar p$,  
the strangeness of $\bar \Lambda$ or $K^+$ and the charge of the 
lepton. One should note that a sample with leptons will be 
enriched in its $\Xi_c^+$ content due to the latter's 
enhanced semileptonic branching ratio, see Eq.(\ref{BRSLEX}).

\section{Numerical estimates}
\label{NUM}

Existing data yield 
\beq 
{\rm BR}(B \to \Lambda _c^+ X) \simeq 6\%
\label{BRL} 
\eeq
Hence one guestimates 
\beq 
{\rm BR}(B \to \Xi _c X) \simeq 1 - 2\%
\label{BRXI}
\eeq
We use as benchmark figure that a sample size of 
500 $fb^{-1}$ will yield about $10^6$ {\em reconstructed} 
$B$ mesons \cite{SHARMA}. With Eqs.(\ref{BRL}) and (\ref{BRXI}) one 
gets about $3 \cdot 10^4$ $\bar p\Lambda_cX$ and 
$10^4$ $\bar p \Xi_c X$ events, where we have assumed equal rates 
for $B \to \bar p +X$ and $B \to \bar n +X$. With 
BR$(\Lambda _c^+ \to p K^-\pi^+) \simeq 5\%$ one estimates 
1500 calibration events $B \to (pK^-\pi^+)_{\Lambda_c} + \bar p +X$.

For 
BR$_{SL}(\Lambda_c) \sim 4\%$ and 
BR$_{SL}(\Xi_c^+) \sim 20\%$
one gets about 1000 $B \to \bar p \Lambda_c X \to \bar p l^+X$ events
and maybe the same number 
$B \to \bar p K\Xi_c^+ X \to \bar p Kl^+X$.

While one will presumably encounter very significant backgrounds, these 
sample sizes should allow meaningful studies of the branching ratios 
and maybe even of the lepton energy {\em spectra}.

\section{Conclusions}
\label{SUM}

Based on the performance demonstrated by BELLE and BABAR we 
can expect to have a million $\Upsilon(4S)$ events in a few years where 
one of the $B$ mesons has been fully reconstructed meaning that all 
remaining tracks belong to the other $B$ meson. This will not only 
revolutionize measurements of $B \to l \nu X_u$, $B \to \gamma X_{s,d}$
in an obvious way, but also allow other measurements that previously 
seemed unfeasible or were not even thought about.

Here we have presented just some examples of this new experimental 
paradigm concerning the measurement of absolute branching ratios 
of charm baryons for exclusive as well as inclusive semileptonic 
modes. Such a sample size should yield about $10^3$ 
identifiable $\Lambda _c \to l^+X$ events with possibly a 
similar number for $\Xi_c^+ \to l^+ X$. If so one can  
contemplate  even to study the lepton spectrum. It could provide us with 
insights into the validity of the HQE and limitations to duality 
at the charm scale beyond what can be learnt from the integrated width.

In studying inclusive semileptonic decays of charm baryons, a 
{\em partial} reconstruction of the first $B$ meson might 
actually suffice; 
in that case considerably larger sample sizes might become 
available. In any case, pilot studies based on partial reconstruction 
should be undertaken well before 500 $fb^{-1}$ have been 
accumulated.

\vspace*{.2cm}

{\bf Acknowledgements:}~~This work was initiated by discussions 
at the Aspen Workshop on Heavy Flavor Dynamics, in particular with 
V. Sharma. I am also grateful to J. LoSecco for valuable comments. 
It has been supported by the 
National Science 
Foundation under grant number PHY00-87419.



\begin{thebibliography}{99}

\bibitem{RIO} 
For a review with references to earlier work, see: I.I. Bigi,
Summary talk at  HQ2K -- "Heavy Quarks at Fixed Target" --, Rio de 
Janeiro, Brazil, October 2000, I. Bediaga, J. Miranda and A. Reis 
(eds.), Frascati Physics Series, Vol. 20, 2000.

\bibitem{BRANCO} 
B. Guberina, R. R\" uckl and J. Trampetic, 
{\em Z. Phys.} {\bf C 33} (1986) 297.

\bibitem{MARBELLA} 
B. Blok and M. Shifman, in: Proceed. of the Third Workshop on the 
Physics at a Tau-Charm Factory, Marbella, Spain, June 1993, 
R. and J. Kirkby (eds.), Editions Frontiere, 1994.


\bibitem{CLEO1}
The CLEO collab., preprint hep-ex/0107040.

\bibitem{FOCUS}
The FOCUS collab., preprint hep-ex/0110002.

\bibitem{VOL1} 
M. Voloshin, {\em Phys. Rev.} {\bf D 61} (2000) 074026.


\bibitem{VADE1}
M. Shifman, in: {\em Boris Joffe Festschrift} "At the 
Frontier of Particle Physics -- Handbook of QCD", M. Shifman (ed.), 
World Scientific, 2001, Vol. 3, p. 1447. 
\bibitem{VADE2} 
I.I. Bigi and N.G. Uraltsev, preprint hep-ph/0106346, to 
appear in Int. J. Mod. Phys. A.


\bibitem{WA}
I.I. Bigi and N.G. Uraltsev, {\em Nucl. Phys.} {\bf B 423} 
(1994) 33; {\em Z. Phys.} {\bf C 62} 
(1994) 623.

\bibitem{VOLOSHIN} 
M. Voloshin, {\em Phys. Lett.} {\bf B 385} (1996) 369.

\bibitem{DDBAR}
I.I. Bigi and N.G. Uraltsev, {\em Nucl. Phys.} {\bf B 592} 
(2001) 92.


\bibitem{CLEO2}
D.E. Jaffe {\em et al.}, CLEO collab., 
{\em Phys. Rev.} {\bf D 62} (2000) 072005.

\bibitem{SHARMA}
V. Sharma, talk given at the Aspen Workshop on Heavy Flavor 
Dynamics, August 2001.


















\end{thebibliography}
\end{document}